\documentstyle[prb,aps,times,amssymb,amsbsy,twocolumn]{revtex}  
\input{epsf.sty} 

\begin{document}    
\title{The temperature dependent bandstructure of a ferromagnetic
  semiconductor film} 
\author{R.~Schiller\cite{email} and W.~Nolting} 
\address{Humboldt-Universit\"at zu Berlin, Institut f\"ur Physik,
       Invalidenstra{\ss}e 110, D-10115 Berlin, Germany} 
\date{\today} 
\maketitle

\begin{abstract}
The electronic quasiparticle spectrum of a ferromagnetic film is
investigated within the framework of the {\em s-f} model. Starting from
the exact solvable case of a single electron in an
otherwise empty conduction band being exchange coupled to a
ferromagnetically saturated localized spin system we extend the theory to
finite temperatures. Our approach is a moment-conserving decoupling
procedure for suitable defined Green functions. The theory for finite
temperatures evolves continuously from the exact limiting case.
The restriction to zero conduction band occupation may be regarded as a
proper model description for ferromagnetic semiconductors like EuO and
EuS.
Evaluating the theory for a simple cubic film cut parallel to the (100) crystal
plane, we find some marked correlation effects which
depend on the spin of the test electron, on the exchange coupling, and
on the temperature of the local-moment system.
\end{abstract}

\section{Introduction} 
\label{sec:intro} 
Since the mid-70's there has been a growing theoretical interest in the
{\em s-f} (or {\em s-d}) model \cite{Nag74,Nol79a}. The
model works for materials which exhibit {\em local-moment magnetism}: 
for magnetic semiconductors like the europium
chalcogenides EuX (X=O, S, Se, Te) \cite{Wac79}
and for metallic local-moment systems such as Gd, Tb, and Dy
\cite{Leg80}. In the local-moment magnets, the electronic and the
magnetic properties are caused by different groups of electrons. 
Whereas the electronic properties like electrical conductivity are borne
by itinerant electrons in rather broad bands, e.g. {\em 6s}, {\em 5d} 
for Gd, the magnetism is due to a strongly localized partially filled 
{\em 4f}-shell. In the case of Gd and Eu compounds the {\em 4f}-shell is 
exactly half-filled and, because of Hund's rules, has its maximal magnetic
moment of $S=\frac{7}{2}$. 

Many characteristics of the local-moment systems may be explained by a
correlation between the localized magnetic states and the itinerant
electrons. In the {\em s-f} model this correlation is represented by an
intra-atomic exchange interaction. The difference between the {\em s-f}
model and the well-known {\em Kondo lattice model} \cite{TSU97} 
is that in the former 
the exchange interaction is ferromagnetic, favouring parallel
alignment of itinerant electrons and local-moments, whereas in latter
it is antiferromagnetic. That is why the {\em s-f} model has been
recently more and more often referred to as the {\em Ferromagnetic Kondo
lattice model}  \cite{Fur94,DYMM+98}. 

The second aspect of this paper is that of reduced dimensionality. Magnetic
phenomena at surfaces and in thin films attract broad attention both
theoretically and experimentally due to the question of phase
transitions and the variation of magnetic and electronic properties in
dimensionally reduced systems \cite{WDW72,Mil84,FF86b,Bin89,WF91,DMIDL97,HN99}.
One of the most remarkable examples of the outstanding magnetic
properties at surfaces is the existence of magnetically ordered surfaces 
at temperatures where the bulk material is paramagnetic. This effect has 
been first documented for Gd(0001) surfaces by Weller {\em et al.}
\cite{WAG+85} and since then been measured by different groups using a 
wide range of experimental techniques \cite{RE86,RR87,TWW+93}. 
In these experiments for the
difference between the Curie temperature at the surface, 
$T_C ({\rm surface})$, and the Curie temperature of bulk Gd,
$T_C ({\rm bulk})$, values between 17K \cite{RE86} and some 60K
\cite{TWW+93} have been reported. Succeeding the results for Gd, Tb also
was found to have a higher surface Curie temperature, relative to the bulk
\cite{RJR88,RJR89}. 

Contrary to the groups cited above, Donath {\em et al.} \cite{DGP96}, using
spin-resolved photoemission did not find any indication for an
enhanced surface Curie temperature of Gd(0001) surfaces. 
Other controversially discussed surface properties of Gd include 
the temperature dependent behaviour of a Gd(0001) surface state
\cite{DGP96,FSK94,WSLM+96} which is supposed to play an important role
in the interplay between electronic structure and magnetism.
A thorough account on the surface magnetism
of the lanthanides has been recently given by Dowben {\em et al.}
\cite{DMIDL97}. 

It is not only the dimensionally reduced Gd which is of interest here,
even bulk Gd is far from being completely understood. In an earlier
study Nolting {\em et al.} \cite{NDB94} have predicted that the a priori
non-magnetic ({\em 5d},{\em 6s})-conduction and valence bands should
exhibit a marked non-uniform magnetic response at different positions
in the Brillouin zone and for different subbands.. Weakly correlated
({\em s}-like) dispersions show a Stoner-like $T$-dependence of the
exchange splitting. On the other hand, stronger correlated 
({\em d}-like) dispersions split below $T_C$ into four branches, two for 
each spin direction. Their $T$-dependence mainly concerns the spectral
weights of the quasiparticle peaks and not so much the energy
positions. Consequently, an exchange caused splitting remains even for
$T>T_c$. This may be the reason for the fact that the experimental
situation is controversial. Kim {\em et al.} \cite{KAE+92} found a
$T$-dependent spin splitting of occupied conduction electron states,
which collapses in a Stoner-like fashion for $T\rightarrow T_C$. From
photoemission experiments, Li {\em et al.} \cite{LZDO92,LZD+92} conclude
that the exchange splitting must be wave-vector dependent, collapsing
for some ${\bf k}$ values, while for other no collapse occurs as a
function of increasing temperature. This fairly complicated temperature
behaviour in the bulk-material must be expected for Gd-films, too.

It is not at all a trivial task to perform an electronic structure
calculation for a ferromagnetic local-moment film in such a manner as to 
realistically incorporate correlation effects. In a previous paper
\cite{SMN97} we proposed a simplified model which allowed us to
exactly calculate the electronic structure of a 
model film in the limiting case of ferromagnetic saturation 
and empty conduction band, $n=0$. This case is applicable to a film of a 
ferromagnetic semiconductor such as EuO, EuS at $T=0$. Its significance
arises from the fact that all relevant correlation effects which are
found or expected to occur at finite band occupations and arbitrary
temperatures \cite{NDB94,NMJR96,NRMJ96}, do already appear in this
rigorously tractable special case \cite{AE82,ND85}.

In this paper we extend the $T=0$ special case to finite
temperatures. As a result we will calculate the electronic
structure of a local-moment film with a single-electron in an otherwise
empty conduction band within the whole temperature range from $T=0$ and
$T=T_c$.

In the next section we present the model and define the corresponding
many-body problem. Subsequently the model is evaluated in two steps, 
in section \ref{subsec:elsub} for the electronic subsystem, and in
section \ref{subsec:lms} for the local-moment system. Section
\ref{sec:results} is devoted to a detailed discussion of the results
obtained for different film thicknesses and various exchange couplings
and temperatures. Comprehensive conclusions with an outlook on the
possible application of the model to real substances and on the
evaluation of temperature dependent surface states in section
\ref{sec:summary} complete the paper.

\section{Theoretical Model} 
\label{sec:theory} 
We investigate a film consisting of $n$ equivalent layers parallel to the
surface of the film. Each lattice site of the film is indicated by a greek
letter $\alpha$, $\beta$, $\gamma$, $\ldots$ denoting the layer index and a
latin letter $i$, $j$, $k$, $\ldots$ numbering the sites within a given layer.
Each layer possesses two-dimensional translational symmetry. Accordingly, 
the thermodynamic average of any site dependent operator $A_{i\alpha}$ depends
only on the layer index $\alpha$:
\begin{equation}
\label{site_ind_ev}
\left\langle A_{i\alpha} \right\rangle \equiv \left\langle 
A_{\alpha} \right\rangle.
\end{equation} 
The complete {\em s-f} model Hamiltonian
\begin{equation}
\label{sf_ham}
{\cal H} = {\cal H}_s + {\cal H}_f + {\cal H}_{sf} 
\end{equation}
consists of three parts. The first
\begin{equation}
\label{h_s}
{\cal H}_s = \sum_{ij \alpha\beta} T^{\alpha\beta}_{ij} c^+_{i\alpha\sigma}
c_{j\beta\sigma}
\end{equation}
describes the itinerant conduction electrons as {\em s}-electrons. 
$c^+_{i\alpha\sigma}$ and $c_{j\beta\sigma}$ are, respectively 
the creation and annihilation operators of an
electron with the spin $\sigma$ at the lattice site ${\bf R}_{i\alpha}$.
$T^{\alpha\beta}_{ij}$ are the hopping integrals.

Each lattice site ${\bf R}_{i\alpha}$ is occupied by a localized magnetic
moment, represented by a spin operator ${\bf S}_{i\alpha}$. 
These localized moments
are exchange coupled expressed by the Heisenberg Hamiltonian:
\begin{equation}
\label{h_f}
{\cal H}_f = - \sum_{ij\alpha\beta} J^{\alpha\beta}_{ij} 
{\bf S}_{i\alpha} {\bf S}_{j\beta},
\end{equation}
where $J^{\alpha\beta}_{ij}$ are the exchange integrals. The problem with 
the simple Heisenberg model in the form (\ref{h_f}) is that due to the
Mermin-Wagner theorem \cite{MW66} there is no solution showing
collective magnetic order at finite temperature, $T>0$. To avoid this
obstacle we have chosen an extended Hamiltonian for the localized
moments,
\begin{equation}
  \label{h_f_ex}
  {\cal H}^*_f = {\cal H}_f + {\cal H}_A = {\cal H}_f - D_0
  \sum_{i\alpha} (S^z_{i\alpha})^2 ,
\end{equation}
which additionally to the Heisenberg Hamiltonian, ${\cal H}_f$
features a single-ion anisotropy term 
${\cal H}_A$. $D_0$ is the according anisotropy constant, which is
typically smaller by some orders of magnitude than the Heisenberg
exchange interaction, $D_0 \ll J^{\alpha\beta}_{ij}$.

The distinguishing feature of the {\em s-f} model is an 
intra-atomic exchange between
the conduction electrons and the localized {\em f}-spins,
\begin{equation}
\label{h_sf_orig}
{\cal H}_{sf} = - \frac{J}{\hslash} 
\sum_{i\alpha} {\bf S}_{i\alpha} \sigma_{i\alpha}
\end{equation}
Here, $J$ is the {\em s-f} exchange interaction and $\sigma_{i\alpha}$ 
is the Pauli
spin operator of the conduction band electrons. For the materials we are
interested in the {\em s-f} coupling is positive ($J>0$). 
In the case where $J<0$ the
model Hamiltonian (\ref{sf_ham}) is that of the so-called {\em Kondo lattice}.
Using the second-quantized form
of $\sigma_{i\alpha}$ and the abbreviations
\begin{equation}
\label{ladderop}
S^{\sigma}_{j\beta} = S^x_{j\beta} + {\rm i}z_{\sigma} S^y_{j\beta};
\quad z_{\uparrow(\downarrow)} = \pm 1,
\end{equation}
the {\em s-f} Hamiltonian can be written as
\begin{equation}
\label{h_sf}
{\cal H}_{sf} =  - \frac{J}{2} \sum_{i\alpha\sigma} \left( z_{\sigma} S^z_{i\alpha}
n_{i\alpha\sigma} + S^{\sigma}_{i\alpha} c^+_{i\alpha-\sigma} c_{i\alpha\sigma}
\right).
\end{equation}
The most decisive part of the {\em s-f} Hamiltonian (\ref{h_sf}) is the 
second term,
which describes spin exchange processes between the conduction electrons
(\ref{h_s}) and the localized moments (\ref{h_f}).

In general, the alignment of the localized moments will be influenced by the
{\em s-f} interaction, which can mediate an indirect interaction (RKKY) via the
occupied conduction band \cite{NRMJ96}.
However, here we are interested in the electronic quasiparticle spectrum of a
ferromagnetic semiconductor according to a simple test electron in an otherwise
empty conduction band. In this case, the localized spin state
cannot be affected by the {\em s-f} interaction. 
Furthermore, one knows from experiment that typical Heisenberg exchange 
integrals are smaller by some orders of magnitudes than their {\em s-f} 
counterparts. In this respect, it seems appropriate to neglect the Heisenberg
exchange integrals for the calculation of the electronic properties of the
system.
Accordingly, the Hamiltonian
(\ref{sf_ham}) can be split into an electronic, ${\cal H}_s + {\cal H}_{sf}$,
and a magnetic part, ${\cal H}^*_f$, which can be solved separately.

\subsection{The electronic subsystem}
\label{subsec:elsub}
Starting from the Hamiltonian of the electronic subsystem,
\begin{equation}
\label{h_*}
{\cal H}^* = {\cal H}_s + {\cal H}_{sf},
\end{equation}
all physical relevant information of the system can be derived 
from the retarded single-electron Green function:
\begin{eqnarray}
  \label{green_gen}
  G^{\alpha\beta}_{ij\sigma} (E) & = & \left\langle\!\left\langle 
  c_{i\alpha\sigma};
  c^+_{j\beta\sigma} \right\rangle\!\right\rangle_E \nonumber \\
  & = & - {\rm i} \int\limits_0^{\infty} dt \:
  {\rm e}^{-\frac{{\rm i}}{\hslash} Et} \left\langle \left[ 
  c_{i\alpha\sigma}(t), c^+_{j\beta\sigma}(0) \right]_+ \right\rangle.
\end{eqnarray}
Here and in what follows $[.,.]_+$ ($[.,.]_-$) is the anticommutator
(commutator). Conform to the two-dimensional
translational symmetry, we perform a Fourier transformation
within the layers of the film,
\begin{equation}
  \label{green_gen_k}
  G^{\alpha\beta}_{{\bf k}\sigma}(E) = \frac{1}{N} \sum_{ij} 
  {\rm e}^{{\rm i}{\bf k}({\bf R}_i - {\bf R}_j)} 
  G^{\alpha\beta}_{ij\sigma}(E),
\end{equation}
where $N$ is the number of sites per layer, ${\bf k}$ is an in-plane wavevector
from the first 2D-Brillouin zone of the layers
and ${\bf R}_i$ represents the in-plane part of the position vector,
\mbox{${\bf R}_{i\alpha} = {\bf R}_i + {\bf r}_{\alpha}$}. From Eq.\
(\ref{green_gen_k}) we get the local spectral density by
\begin{equation}
  \label{spectral}
  S^{\alpha\beta}_{{\bf k}\sigma}(E) = - \frac{1}{\pi} {\rm Im} 
  G^{\alpha\beta}_{{\bf k}\sigma} (E+{\rm i}0^+),
\end{equation}
which is directly related to observable quantities within angle and spin
resolved direct and inverse photoemission experiments. Finally, the 
wave-vector summation of $S^{\alpha\beta}_{{\bf k}\sigma}(E)$ yields the
layer-dependent (local) quasiparticle density of states:
\begin{equation}
\label{ldos}
\rho_{\alpha\sigma}(E) = \frac{1}{\hslash N} \sum_{{\bf k}} 
S^{\alpha\alpha}_{{\bf k}\sigma}(E).
\end{equation} 
In the following discussion all results will be interpreted in terms of the
spectral density (\ref{spectral}) and the local density of states (\ref{ldos}).

For the solution of the many-body problem posed by Eq.\ (\ref{h_*}) 
we write down the equation of motion of the single-electron Green function
(\ref{green_gen})
\begin{equation}
\label{eom_gen}
E\,G^{\alpha\beta}_{ij\sigma} = \hslash \delta^{\alpha\beta}_{ij} +
\sum_{m\mu} T^{\alpha\mu}_{im} G^{\mu\beta}_{mj\sigma} + 
\langle\!\langle [c_{i\alpha\sigma},{\cal H}_{sf}]_-;
c^+_{j\beta\sigma} \rangle\!\rangle_E,
\end{equation}
where $\delta^{\alpha\beta}_{ij} \equiv \delta_{\alpha\beta} \delta_{ij}$.
The formal solution of Eq.\ (\ref{eom_gen}) can be found by introducing
the self-energy $M^{\alpha\beta}_{ij\sigma}(E)$,
\begin{equation}
  \label{self_gen}
  \left\langle\!\left\langle \left[c_{i\alpha\sigma},{\cal H}_{sf}\right]_-;
  c^+_{j\beta\sigma} \right\rangle\!\right\rangle_E =
  \sum_{m\mu} M^{\alpha\mu}_{im\sigma}(E) G^{\mu\beta}_{mj\sigma}(E),  
\end{equation}
which contains all information about the correlations between the conduction
band and localized moments. After combining Eqs.\ (\ref{eom_gen}) and 
(\ref{self_gen}) and performing a two-dimensional Fourier transform we
see that the formal solution of Eq.\ (\ref{eom_gen}) is given by
\begin{equation}
  \label{form_sol}
  {\bf G}_{{\bf k}\sigma}(E) = \hslash \left(E\,{\bf I}-{\bf T}_{\bf k} -
  {\bf M}_{{\bf k}\sigma}(E) \right)^{-1},
\end{equation}
where ${\bf I}$ represents the ($n\times n$) identity matrix and where
the matrices ${\bf G}_{{\bf k}\sigma}(E)$, ${\bf T}_{\bf k}$, and 
${\bf M}_{{\bf k}\sigma}(E)$
have as elements the layer-dependent functions 
$G^{\alpha\beta}_{{\bf k}\sigma}(E)$, $T^{\alpha\beta}_{{\bf k}}$, and 
$M^{\alpha\beta}_{{\bf k}\sigma}(E)$, respectively.

To explicitly get the self-energy in Eq.\ (\ref{self_gen}) we evaluate
the Green function
\begin{equation}
  \label{c_hsf}
  \left\langle\!\left\langle \left[c_{i\alpha\sigma},{\cal H}_{sf}\right]_-;
  c^+_{j\beta\sigma} \right\rangle\!\right\rangle_E =
  - \frac{J}{2} \left( z_{\sigma} {\it \Gamma}^{\alpha\alpha\beta}_{iij\sigma}
  + F^{\alpha\alpha\beta}_{iij\sigma} \right).
\end{equation}
Here the two higher Green functions,
\begin{eqnarray}
  \label{gamma}
  {\it \Gamma}^{\alpha\gamma\beta}_{ikj\sigma}(E) & = &
  \left\langle\!\left\langle S^z_{i\alpha} c_{k\gamma\sigma} ;
  c^+_{j\beta\sigma} \right\rangle\!\right\rangle_E ,\\
  \label{spin-flip}
  F^{\alpha\alpha\beta}_{iij\sigma}(E) & = &
  \left\langle\!\left\langle S^{-\sigma}_{i\alpha} c_{k\gamma-\sigma} ;
    c^+_{j\beta\sigma} \right\rangle\!\right\rangle_E ,
\end{eqnarray}
originate form the two terms of the
{\em s-f} Hamiltonian (\ref{h_sf}) and will be referred  to as the {\em Ising} and
the {\em Spin-flip} function, respectively. Considering the equations
of motion for these two Green functions we encounter the two higher
Green functions 
$\langle\!\langle [S^z_{i\alpha}  c_{k\gamma\sigma}, {\cal H}_{sf}]_- ;
c^+_{j\beta\sigma} \rangle\!\rangle_E$ 
and 
$\langle\!\langle  [S^{-\sigma}_{i\alpha} c_{k\gamma-\sigma}, {\cal
  H}_{sf} ]_- ; c^+_{j\beta\sigma} \rangle\!\rangle_E$.
Since we consider an empty conduction band the
thermodynamic average in the Green functions has to be computed with the 
electron vacuum state $\left| n=0 \right\rangle$. 
From the definition of the {\em s-f} Hamiltonian (\ref{h_sf}) we then see that 
$\left\langle n=0\right|{\cal H}_{sf}=0$ and, accordingly,
\begin{eqnarray*}
  \left\langle\!\left\langle \left[S^z_{i\alpha},{\cal H}_{sf}\right]_-
  c_{k\gamma\sigma};c^+_{j\beta\sigma} \right\rangle\!\right\rangle_E
  & \stackrel{n\rightarrow 0}{\longrightarrow} & 0 , \\
  \left\langle\!\left\langle \left[S^{-\sigma}_{i\alpha},{\cal H}_{sf}\right]_-
  c_{k\gamma-\sigma};c^+_{j\beta\sigma} \right\rangle\!\right\rangle_E
  & \stackrel{n\rightarrow 0}{\longrightarrow} & 0 .
\end{eqnarray*}
Hence, for the equations of motion of the Ising and the Spin-flip
function we get
\begin{eqnarray}
  \label{eom_ga_gen}
  \lefteqn{\sum_{m\mu} \left(E\delta^{\gamma\mu}_{km}-T^{\gamma\mu}_{km} 
  \right) {\it \Gamma}^{\alpha\mu\beta}_{imj\sigma}(E)} \nonumber \\ & = &
  \hslash \left\langle S^z_{\alpha} \right\rangle
  \delta^{\gamma\beta}_{kj}
  + \left\langle\!\left\langle S^z_{i\alpha} \left[ c_{k\gamma\sigma},
  {\cal H}_{sf} \right]_- ; c^+_{j\beta\sigma}
  \right\rangle\!\right\rangle_E \\[2ex]
  \lefteqn{\label{eom_f_gen}
  \sum_{m\mu} \left(E\delta^{\gamma\mu}_{km}-T^{\gamma\mu}_{km} \right)
  F^{\alpha\mu\beta}_{imj\sigma}(E)} \nonumber \\ & = &
  \left\langle\!\left\langle S^{-\sigma}_{i\alpha} \left[ c_{k\gamma-\sigma},
  {\cal H}_{sf} \right]_- ; c^+_{j\beta\sigma}
  \right\rangle\!\right\rangle_E .
\end{eqnarray}
On the right-hand side of these equations appear further higher Green
functions which prevent a direct solution and require an approximative 
treatment. The treatment is different for the non-diagonal terms, 
$(i,\alpha)\ne(k,\gamma)$ and for the diagonal terms, 
$(i,\alpha)=(k,\gamma)$. In the first case we use a self-consistent
so-called {\em self-energy approach} which results in a decoupling of
the equations of motion.
For the diagonal terms, $(i,\alpha)=(k,\gamma)$, this approach is
replaced by a moment technique which takes the local correlations better 
into account.

\paragraph{Non-diagonal terms $(i,\alpha)\ne(k,\gamma)$:}

The definition of the self-energy (\ref{self_gen}) formally corresponds
to the substitution
\begin{equation}
  \label{subst_se}
  \left[c_{i\alpha\sigma},{\cal H}_{sf}\right]_- \longrightarrow
  \sum_{m\mu} M^{\alpha\mu}_{im\sigma}(E) c_{m\mu\sigma}
\end{equation}
within the brackets of the Green function. The inspection of the
spectral decomposition of the two functions in Eq.\ (\ref{self_gen})
reveals that both, 
$\langle\!\langle [c_{i\alpha\sigma},{\cal H}_{sf}]_-;
c^+_{j\beta\sigma} \rangle\!\rangle_E$ and
$\langle\!\langle c_{i\alpha\sigma}; c^+_{j\beta\sigma}
\rangle\!\rangle_E$,
have the same pole structure and can differ only by the spectral weights 
of their poles. The equality of both sides in Eq.\ (\ref{self_gen}) is
installed by the self-energy components $M^{\alpha\beta}_{ij\sigma}(E)$.
Inspecting now the spectral representations of the two Green functions 
$\langle\!\langle S^{-\sigma}_{i\alpha} [c_{k\gamma\sigma},{\cal
  H}_{sf}]_-;c^+_{j\beta\sigma} \rangle\!\rangle_E$
and 
$\langle\!\langle S^{-\sigma}_{i\alpha} c_{k\gamma\sigma};c^+_{j\beta\sigma}
\rangle\!\rangle_E$ 
we notice that the additional spin operator
$S^{-\sigma}_{i\alpha}$ selects for both only those poles of the original Green
functions without spin operator which are connected with a
spin-flip of the electron. Hence, the poles of
these two functions build subset of the poles of the two 
Green functions from Eq.\ (\ref{self_gen}) and are, therefore, identical
to each other.
Again, only the weights of the poles can differ. In analogy to
Eqs.\ (\ref{self_gen}) and (\ref{subst_se}) we now propose to use the
plausible ansatz 
\begin{eqnarray}
  \label{sea1}
  \lefteqn{\left\langle\!\left\langle S^{-\sigma}_{i\alpha} 
  \left[c_{k\gamma-\sigma},{\cal H}_{sf}\right]_-;c^+_{j\beta\sigma} 
  \right\rangle\!\right\rangle_E} \nonumber \\
  & & \approx \sum_{m\mu} M^{\gamma\mu}_{km-\sigma}(E)
  \left\langle\!\left\langle S^{-\sigma}_{i\alpha} c_{m\mu-\sigma} ;
  c^+_{j\beta\sigma} \right\rangle\!\right\rangle_E.
\end{eqnarray}
A similar reasoning can be used for:
\begin{eqnarray}
  \label{sea2}
  \lefteqn{\left\langle\!\left\langle S^z_{i\alpha} 
  \left[c_{k\gamma\sigma},{\cal H}_{sf}\right]_-;c^+_{j\beta\sigma} 
  \right\rangle\!\right\rangle_E} \nonumber \\
  & & \approx \sum_{m\mu} M^{\gamma\mu}_{km\sigma}(E)
  \left\langle\!\left\langle S^z_{i\alpha} c_{m\mu\sigma} ;
  c^+_{j\beta\sigma} \right\rangle\!\right\rangle_E, 
\end{eqnarray}
with the difference that here the additional spin operator
$S^z_{i\alpha}$ does not change the original pole structure but
merely modifies the spectral weights of the poles.
On the right-hand sides of Eqs.\ (\ref{sea1}) and (\ref{sea2}) we find
the already known Spin-flip and Ising function, respectively. Hence, for 
$(i,\alpha)\ne(k,\gamma)$, the Eqs.\ (\ref{eom_gen}), (\ref{c_hsf}),
(\ref{eom_ga_gen}), (\ref{eom_f_gen}), (\ref{sea1}), and (\ref{sea2})
build a closed system .

\paragraph{Diagonal elements $(i,\alpha)=(k,\gamma)$:}

We start with the explicit evaluation of the higher Green functions on
the right-hand sides of Eqs.\ (\ref{eom_ga_gen}) and
(\ref{eom_f_gen}). For Eq.\ (\ref{eom_f_gen}) we get, for
$(i,\alpha)=(k,\gamma)$,
\begin{eqnarray}
  \label{eq_high_f}
  \lefteqn{\left\langle\!\left\langle S^{-\sigma}_{i\alpha} \left[
  c_{i\alpha-\sigma} , {\cal H}_{sf} \right]_-; c^+_{j\beta\sigma}
  \right\rangle\!\right\rangle_E} \nonumber \\
  & & = \frac{J}{2} \left( z_{\sigma}
    \dot{F}^{\alpha\alpha\beta}_{iij\sigma}(E) - 
    \ddot{F}^{\alpha\alpha\beta}_{iij\sigma}(E) \right),
\end{eqnarray}
where we have abbreviated
\begin{mathletters}
  \label{def_high_f}
  \begin{eqnarray}
    \label{def_high_f_1}
    \dot{F}^{\alpha\alpha\beta}_{iij\sigma}(E) & = &
    \left\langle\!\left\langle S^{-\sigma}_{i\alpha} S^z_{i\alpha}
    c_{i\alpha-\sigma} ; c^+_{j\beta\sigma}
    \right\rangle\!\right\rangle_E, \\
    \ddot{F}^{\alpha\alpha\beta}_{iij\sigma}(E) & = &
    \left\langle\!\left\langle S^{-\sigma}_{i\alpha} S^{\sigma}_{i\alpha}
    c_{i\alpha\sigma} ; c^+_{j\beta\sigma}
    \right\rangle\!\right\rangle_E.
  \end{eqnarray}
\end{mathletters}
The analogous evaluation of the higher Green function in Eq.\
(\ref{eom_ga_gen}) does not require any further higher Green functions,
because it can be expressed in terms of already known Green functions:
\begin{eqnarray}
  \label{eq_high_ga}
  \lefteqn{\!\!\langle\!\langle S^z_{i\alpha} [ c_{i\alpha\sigma} ,
        {\cal H}_{sf} ]_- ; c^+_{j\beta\sigma}
    \rangle\!\rangle_E \! + \! z_{\sigma}
  \langle\!\langle S^{-\sigma}_{i\alpha} [
  c_{i\alpha-\sigma} , {\cal H}_{sf} ]_-; c^+_{j\beta\sigma}
  \rangle\!\rangle_E} \nonumber \\
  & = & \frac{J\hslash}{2} ( 
  {\it \Gamma}^{\alpha\alpha\beta}_{iij\sigma}(E) + z_{\sigma}
  F^{\alpha\alpha\beta}_{iij\sigma}(E) - z_{\sigma}\hslash S(S+1)
  G^{\alpha\beta}_{ij\sigma}(E) ). \nonumber \\
\end{eqnarray}
As Eq.\ (\ref{eq_high_f}), the above relation is still exact. To get a close
system of equations we are left with the determination of the functions
$\dot{F}^{\alpha\alpha\beta}_{iij\sigma}(E)$ and
$\ddot{F}^{\alpha\alpha\beta}_{iij\sigma}(E)$. Both fulfil exact
relations which will be used to derive satisfying approximations. For
spin $S=\frac{1}{2}$ we find for all temperatures:
\begin{mathletters}
  \label{lc_s_onehalf}
  \begin{eqnarray}
    \label{lc_s_onehalf_1}
    \dot{F}^{\alpha\alpha\beta}_{iij\sigma}(E) \Big|_{S=\frac{1}{2}} & = & 
    {\textstyle \frac{1}{2}}
    z_{\sigma} \hslash F^{\alpha\alpha\beta}_{iij\sigma}(E),\\
    \label{lc_s_onehalf_2}
    \ddot{F}^{\alpha\alpha\beta}_{iij\sigma}(E) \Big|_{S=\frac{1}{2}} & = & 
    {\textstyle \frac{1}{2}} \hslash^2 G^{\alpha\beta}_{ij\sigma}(E) -
    z_{\sigma} \hslash {\it \Gamma}^{\alpha\alpha\beta}_{iij\sigma}(E).
  \end{eqnarray}
\end{mathletters}
On the other hand, in the case of ferromagnetic saturation,
$\langle S^z_{\alpha}\rangle \equiv S$, it holds for arbitrary spin:
\begin{mathletters}
  \label{lc_t_zero}
  \begin{eqnarray}
    \label{lc_t_zero_1}
    \dot{F}^{\alpha\alpha\beta}_{iij\sigma}(E) \Big|_{T=0} & = & 
    \hslash \left( (S-{\textstyle \frac{1}{2}}) + 
    {\textstyle \frac{1}{2}}z_{\sigma} \right) 
    F^{\alpha\alpha\beta}_{iij\sigma}(E),\\
    \label{lc_t_zero_2}
    \ddot{F}^{\alpha\alpha\beta}_{iij\sigma}(E) \Big|_{T=0} & = & 
    \hslash^2 S G^{\alpha\beta}_{ij\sigma}(E) - z_{\sigma}
    \hslash {\it \Gamma}^{\alpha\alpha\beta}_{iij\sigma}(E).
  \end{eqnarray}
\end{mathletters}
The exact limiting cases (\ref{lc_s_onehalf}) and (\ref{lc_t_zero})
suggest the general structures:
\begin{mathletters}
  \label{gen_struc}
  \begin{eqnarray}
    \label{gen_struc_1}
    \dot{F}^{\alpha\alpha\beta}_{iij\sigma}(E) & = & 
    \kappa^{(1)}_{\alpha\sigma} G^{\alpha\beta}_{ij\sigma}(E) +
    \lambda^{(1)}_{\alpha\sigma} F^{\alpha\alpha\beta}_{iij\sigma}(E), \\
    \label{gen_struc_2}
    \ddot{F}^{\alpha\alpha\beta}_{iij\sigma}(E) & = & 
    \kappa^{(2)}_{\alpha\sigma} G^{\alpha\beta}_{ij\sigma}(E) +
    \lambda^{(2)}_{\alpha\sigma} 
    {\it \Gamma}^{\alpha\alpha\beta}_{iij\sigma}(E).
  \end{eqnarray}
\end{mathletters}
For the five Green functions of the type 
$\langle\!\langle A;B \rangle\!\rangle_E$ in Eqs.\ (\ref{gen_struc}) we
can calculate the spectral moments,
\begin{equation}
  \label{def_specmom}
   M^{(n)}_{AB} = \left\langle \left( {\rm i} \hslash
   \frac{\partial}{\partial t} \right)^n \left[
   A(t) , B(0) \right]_+ \right\rangle_{t=0} ,
\end{equation}
where $n=1,2,\ldots$ . Because of the equivalent relation
\begin{equation}
  \label{specmom_green}
  M^{(n)}_{AB} = - \frac{1}{\pi\hslash} \int\limits_{-\infty}^{\infty}
  dE \: E^n {\rm Im} \left\langle\!\left\langle A;B
    \right\rangle\!\right\rangle_E ,
\end{equation}
the moments can be used to fix the coefficients
$\kappa^{(m)}_{\alpha\sigma}$ and $\lambda^{(m)}_{\alpha\sigma}$ in
Eqs.\ (\ref{gen_struc}). After tedious but straightforward calculations, 
we get
\begin{eqnarray}
  \label{coeff} 
  & \kappa^{(1)}_{\alpha\sigma}=0, \quad \kappa^{(2)}_{\alpha\sigma} =
  \langle S^{-\sigma}_{\alpha} S^{\sigma}_{\alpha} \rangle
  - \lambda^{(2)}_{\alpha\sigma} \langle S^z_{\alpha} \rangle, & 
  \nonumber \\
  & \lambda^{(1)}_{\alpha\sigma} = 
    \displaystyle \frac{\langle S^{-\sigma}_{\alpha}
    S^{\sigma}_{\alpha} S^z_{\alpha} \rangle + z_{\sigma}
    \langle S^{-\sigma}_{\alpha} S^{\sigma}_{\alpha} \rangle}
    {\langle S^{-\sigma}_{\alpha} S^{\sigma}_{\alpha} \rangle}, &
  \\
  & \lambda^{(2)}_{\alpha\sigma} =
    \displaystyle \frac{\langle S^{-\sigma}_{\alpha}
    S^{\sigma}_{\alpha} S^z_{\alpha} \rangle - \langle S^z_{\alpha}
    \rangle \langle S^{-\sigma}_{\alpha} S^{\sigma}_{\alpha} \rangle}
    {\langle (S^z_{\alpha})^2 \rangle - \langle S^z_{\alpha} \rangle^2}.
    & \nonumber
\end{eqnarray}
The coefficients are determined by {\em f}-spin correlation functions,
which will be determined at a later stage. 

The Eqs.\ (\ref{eom_gen}), (\ref{c_hsf}), (\ref{eom_ga_gen}),
(\ref{eom_f_gen}), (\ref{sea1})--(\ref{eq_high_f}), (\ref{eq_high_ga}),
(\ref{gen_struc}), and (\ref{coeff}) represent a closed system, which
can be solved self-consistently. Before proceeding we assume that the
self-energy from Eq.\ (\ref{self_gen}) is a local entity
\begin{equation}
  \label{local_self}
  M^{\alpha\beta}_{{\bf k}\sigma}(E) \equiv \delta_{\alpha\beta} 
  M^{\alpha}_{\sigma}(E).
\end{equation}
The reason for the ${\bf k}$ independence can be traced back to the
neglect of magnon energies \cite{NMJR96}. The restriction to the
diagonal elements in the greek indices denoting the layers is in that
sense the transfer of the ${\bf k}$ independence in the case of three
dimensions \cite{NMJR96} to the film geometries discussed in this
paper. Furthermore one can show that the assumption (\ref{local_self}) is 
not necessary for the following calculations but merely drastically 
simplifies them.

We can now use Eqs.\
(\ref{sea1})--(\ref{eq_high_f}), (\ref{eq_high_ga}),
(\ref{gen_struc}), and (\ref{coeff}) to evaluate the Ising and the
Spin-flip functions in Eqs.\ (\ref{eom_ga_gen}) and
(\ref{eom_f_gen}). As the result we get the Fourier transformed Ising
and Spin-flip functions. According to Eq.\ (\ref{c_hsf}) we can restrict 
our attention to the diagonal elements 
${\it \Gamma}^{\alpha\alpha\beta}_{{\bf kq}\sigma}(E)$ and 
$F^{\alpha\alpha\beta}_{{\bf kq}\sigma}(E)$. 
After subsequent ${\bf q}$-summation we eventually get, using Eqs.\
(\ref{form_sol}) and (\ref{local_self}):
\begin{mathletters}
  \label{eq_sum}
  \begin{eqnarray}
    \label{eq_sum_ga}
    \hslash \sum_{{\bf q}} 
    {\it \Gamma}^{\alpha\alpha\beta}_{{\bf kq}\sigma} & = &
    \sqrt{N} \hslash \langle S^z_{\alpha} \rangle 
    G^{\alpha\beta}_{{\bf k}\sigma} - G^{\alpha\alpha}_{{\bf 0}\sigma}
    \Big\{ M^{\alpha}_{\sigma} \sum_{{\bf q}} 
    {\it \Gamma}^{\alpha\alpha\beta}_{{\bf kq}\sigma} \nonumber \\
    & & - \frac{J}{2} \Big( z_{\sigma} (\kappa^{(2)}_{\alpha\sigma} -
    \hslash^2 S (S+1)) \sqrt{N} G^{\alpha\beta}_{{\bf k}\sigma} 
    \nonumber \\
    & & + (z_{\sigma}\hslash - \lambda^{(1)}_{\alpha\sigma})
    \sum_{{\bf q}} F^{\alpha\alpha\beta}_{{\bf kq}\sigma} \nonumber \\
    & & + (\hslash + z_{\sigma} \lambda^{(2)}_{\alpha\sigma} )
    \sum_{{\bf q}} {\it \Gamma}^{\alpha\alpha\beta}_{{\bf kq}\sigma}
    \Big) \Big\}, \\
    \label{eq_sum_f}
    \hslash \sum_{{\bf q}} F^{\alpha\alpha\beta}_{{\bf kq}\sigma} & = &
    - G^{\alpha\alpha}_{{\bf 0}-\sigma} \Big\{ M^{\alpha}_{-\sigma}
    \sum_{{\bf q}} F^{\alpha\alpha\beta}_{{\bf kq}\sigma} + \frac{J}{2}
    \Big( \kappa^{(2)}_{\alpha\sigma} \sqrt{N} 
    G^{\alpha\beta}_{{\bf k}\sigma} \nonumber \\
    & & - z_{\sigma}
    \lambda^{(1)}_{\alpha\sigma} \sum_{{\bf q}}
    F^{\alpha\alpha\beta}_{{\bf kq}\sigma} +
    \lambda^{(2)}_{\alpha\sigma} \sum_{{\bf q}} 
    {\it \Gamma}^{\alpha\alpha\beta}_{{\bf kq}\sigma} \Big) \Big\},
  \end{eqnarray}
\end{mathletters}
where we have introduced
\begin{equation}
  \label{def_g0}
  G^{\alpha\alpha}_{{\bf 0}\sigma}(E) = \frac{1}{N} \sum_{{\bf k}}
  G^{\alpha\alpha}_{{\bf k}\sigma}(E).
\end{equation}
The set of equations (\ref{eq_sum}) can be solved to express the sums 
$\sum_{{\bf q}}{\it \Gamma}^{\alpha\alpha\beta}_{{\bf kq}\sigma}(E)$ and 
$\sum_{{\bf q}}F^{\alpha\alpha\beta}_{{\bf kq}\sigma}(E)$ in terms of
the single-electron Green function $G^{\alpha\beta}_{{\bf k}\sigma}(E)$.
However, by inspecting Eqs.\ (\ref{eq_sum}) we see that these
expressions will still contain the layer and spin-dependent self-energy
$M^{\alpha}_{\sigma}(E)$. 

To solve this problem we combine Eqs.\ (\ref{self_gen}) and (\ref{c_hsf})
and get, after Fourier transformation:
\begin{equation}
  \label{self_ga_f}
  M^{\alpha}_{\sigma} G^{\alpha\beta}_{{\bf k}\sigma}
  = - \frac{J}{2 \sqrt{N}} \left( z_{\sigma} \sum_{{\bf q}}
  {\it \Gamma}^{\alpha\alpha\beta}_{{\bf kq}\sigma} + \sum_{{\bf q}}
  F^{\alpha\alpha\beta}_{{\bf kq}\sigma} \right).
\end{equation}
Combining this equation with the results obtained for 
$\sum_{{\bf q}}{\it \Gamma}^{\alpha\alpha\beta}_{{\bf kq}\sigma}(E)$ and 
$\sum_{{\bf q}}F^{\alpha\alpha\beta}_{{\bf kq}\sigma}(E)$
from Eqs. (\ref{eq_sum}) we eventually get an implicit set of equations
for the layer and spin-dependent electronic self-energy:
\begin{equation}
  \label{small_self}
  M^{\alpha}_{\sigma}(E) = -\frac{J}{2} m^{\alpha}_{\sigma} (E), \quad
  m^{\alpha}_{\sigma} (E) =
  \frac{Z^{\alpha}_{\sigma}(E)}{N^{\alpha}_{\sigma}(E)},
\end{equation}
where the numerator and the denominator, respectively, are given by
\begin{mathletters}
  \label{num_den}
  \begin{eqnarray}
    \label{num_den_z}
    Z^{\alpha}_{\sigma} & = & z_{\sigma} \hslash^2 \langle S^z_{\alpha}
    \rangle + \frac{J}{2} \Big\{ \big( \kappa^{(2)}_{\alpha\sigma} -
    \hslash^2 S(S+1) \big) G^{\alpha\alpha}_{{\bf 0}\sigma}
    \nonumber \\ & &
    - \big( (\lambda^{(1)}_{\alpha\sigma} +
    \lambda^{(2)}_{\alpha\sigma} + z_{\sigma} m^{\alpha}_{-\sigma} )
    \hslash \langle S^z_{\alpha} \rangle + \hslash
    \kappa^{(2)}_{\alpha\sigma} \big) G^{\alpha\alpha}_{{\bf 0}-\sigma}
     \Big\}
    \nonumber \\ & & + \frac{J^2}{4} \Big\{
    z_{\sigma} \hslash^2 S(S+1)
    (\lambda^{(1)}_{\alpha\sigma} + \lambda^{(2)}_{\alpha\sigma}
    + z_{\sigma} m^{\alpha}_{-\sigma}) 
    \nonumber \\ & & 
    + \kappa^{(2)}_{\alpha\sigma}
    (m^{\alpha}_{\sigma} - m^{\alpha}_{-\sigma}) \Big\} 
    G^{\alpha\alpha}_{{\bf 0}\sigma} G^{\alpha\alpha}_{{\bf 0}-\sigma} ,\\
    \label{num_den_n}
    N^{\alpha}_{\sigma} & = & \hslash^2 - \frac{J}{2} \Big\{ (\hslash +
    z_{\sigma} \lambda^{(2)}_{\alpha\sigma} + m^{\alpha}_{\sigma})
    G^{\alpha\alpha}_{{\bf 0}\sigma} + (z_{\sigma}
    \lambda^{(1)}_{\alpha\sigma} 
    \nonumber \\ & &
    + m^{\alpha}_{-\sigma})
    G^{\alpha\alpha}_{{\bf 0}-\sigma} \Big\} 
    + \frac{J^2}{4} \Big\{ (m^{\alpha}_{\sigma} + \hslash)
    (m^{\alpha}_{-\sigma} + z_{\sigma} \lambda^{(1)}_{\alpha\sigma})
    \nonumber \\ & &
    + z_{\sigma} \lambda^{(2)}_{\alpha\sigma} (m^{\alpha}_{-\sigma} +
    \hslash) \Big\} G^{\alpha\alpha}_{{\bf 0}\sigma} 
    G^{\alpha\alpha}_{{\bf 0}-\sigma}.
  \end{eqnarray}
\end{mathletters}

The implicit set of equations (\ref{small_self}) and (\ref{num_den}) now
enables us to self-consistently evaluate the self-energy of the system
provided that the {\em f}-spin correlation functions from Eqs.\
(\ref{coeff}) are known. These will be evaluated in the next section.

\subsection{The local-moment system}
\label{subsec:lms}
The system of localized {\em f}-moments is described by the extended 
Heisenberg Hamiltonian (\ref{h_f_ex}) which we write down again for
convenience:
\begin{equation}
  \label{h_f_new}
  {\cal H}^*_f = - \sum_{ij\alpha\beta} J^{\alpha\beta}_{ij} 
  {\bf S}_{i\alpha} {\bf S}_{j\beta} - D_0
  \sum_{i\alpha} (S^z_{i\alpha})^2.
\end{equation}
Here we want to stress once more that the single-ion anisotropy
constant $D_0$ is small compared to the Heisenberg exchange interaction,
$D_0 \ll J^{\alpha\beta}_{ij}$. By defining the magnon Green function
\begin{equation}
  \label{green_heis}
  D^{\alpha\beta}_{ij}(E) = \left\langle\!\left\langle
  S^+_{i\alpha}; S^-_{j\beta} \right\rangle\!\right\rangle_E,
\end{equation}
we can calculate the {\em f}-spin correlation functions by evaluating
the equation of motion
\begin{equation}
  \label{eom_green_heis}
  E\,D^{\alpha\beta}_{ij}(E) = 2 \hslash^2 \delta_{\alpha\beta} \langle
  S^z_{\alpha} \rangle + \left\langle\!\left\langle \left[
        S^+_{i\alpha}, {\cal H}^*_f \right]_- ; S^-_{j\beta}
        \right\rangle\!\right\rangle_E .
\end{equation}
The evaluation of this equation of motion involves the decoupling of the
higher Green functions on its right-hand side,
originating from the Heisenberg term. ${\cal H}_f$,
and the anisotropy term, ${\cal H}_A$, using the Random Phase
Approximation (RPA) and a decoupling proposed by Lines \cite{Lin67},
respectively. 
The details of the calculation can be found in a
previous paper \cite{SN99}. For brevity we restrict ourselves to 
present here only the results. 
For the layer-dependent magnetizations of the {\em f}-spin
system we get
\begin{equation}
  \label{s_z}
  \frac{\langle S^z_{\alpha}\rangle}{\hslash}  = 
  \frac{(1+\varphi_{\alpha})^{2S+1} (S-\varphi_{\alpha}) +
    \varphi^{2S+1}_{\alpha} (S+1+\varphi_{\alpha})}
    {\varphi^{2S+1}_{\alpha} - (1+\varphi_{\alpha})^{2S+1}},
\end{equation}
where
\begin{equation}
  \label{varphi}
  \varphi_{\alpha} = \frac{1}{N} \sum_{{\bf k}} \sum_{\gamma}
  \frac{\chi_{\alpha\alpha\gamma}({\bf k})}
  {{\rm e}^{\beta E_{\gamma}({\bf k})} - 1},
\end{equation}
where, again, $N$ is the number of sites per layer and
$\beta=\frac{1}{k_{{\rm B}}T}$.
The summation $\sum_{\gamma}$ in Eq.\ (\ref{varphi}) runs over the $n$
poles $E_{\gamma}({\bf k})$ of the Green function
$D^{\alpha\beta}_{{\bf k}}(E)$ and the 
$\chi_{\alpha\alpha\gamma}({\bf k})$ is the weight of the $\gamma$'th
pole in the diagonal element of the Green function 
$D^{\alpha\alpha}_{{\bf k}}(E)$. The poles and the weights can be
calculated from the solution of Eq.\ (\ref{eom_green_heis}):
\begin{equation}
  \label{sol_eom_heis}
  D^{\alpha\beta}_{{\bf k}}(E) = 2 \hslash^2
  \left( \begin{array}{ccc}\langle S^z_1 \rangle & & 0 \\
  & \ddots & \\ 0 & & \langle S^z_n \rangle \end{array} \right)
  \cdot (E\, {\bf I} - {\bf A})^{-1},
\end{equation}
with
\begin{equation}
  \label{a_al_be}
  \frac{({\bf A})^{\alpha\beta}}{\hslash} = \big( D_0 \Phi_{\alpha} +
  2 \sum_{\gamma} J^{\alpha\gamma}_{{\bf 0}} \langle S^z_{\gamma}
  \rangle \big) \delta_{\alpha\beta} - 2 J^{\alpha\beta}_{{\bf k}}
  \langle S^z_{\alpha} \rangle.
\end{equation}
The $\Phi_{\alpha}$ come from the decoupling of the higher Green
function on the right-hand side of Eq.\ (\ref{eom_green_heis}) which
originates from the anisotropy Hamiltonian ${\cal H}_{A}$ according to
Lines \cite{Lin67,SN99} and are given by:
\begin{equation}
  \label{lines}
  \Phi_{\alpha} = \frac{2\hslash^2 S(S+1) - 3 \hslash \langle
    S^z_{\alpha} \rangle (1+2 \varphi_{\alpha})}{\langle S^z_{\alpha} \rangle}.
\end{equation}

Having obtained the layer-dependent magnetizations (\ref{s_z}) and the
$\varphi_{\alpha}$ we can now express all the other {\em f}-spin
correlation functions appearing in Eqs. (\ref{coeff}) via the relations
\begin{mathletters}
  \label{rel_coeff}
  \begin{eqnarray}
    \label{rel_coeff_s-s+}
    \langle S^-_{\alpha} S^+_{\alpha} \rangle & = & 
    2 \hslash \langle S^z_{\alpha} \rangle \varphi_{\alpha}, \\
    \langle (S^z_{\alpha})^2 \rangle & = & \hslash^2 S (S+1)
    \langle S^z_{\alpha} \rangle (1+2\varphi_{\alpha}),\\
    \langle (S^z_{\alpha})^3 \rangle & = & \hslash^3 S(S+1)
    \varphi_{\alpha} + \hslash^2 \langle S^z_{\alpha} \rangle \big(
    S(S+1) + \varphi_{\alpha} \big) \nonumber \\
    & & - \hslash \langle (S^z_{\alpha})^2 \rangle (1+2\varphi_{\alpha}),
  \end{eqnarray}
\end{mathletters}
and the general spin-operator equality
\begin{equation}
  \label{spin_op_rel}
  S^{\sigma}_{i\alpha} S^{-\sigma}_{i\alpha} = \hslash^2 S(S+1) +
  z_{\sigma} \hslash S^z_{i\alpha} - (S^z_{i\alpha})^2 .
\end{equation}

Mediated by Eqs.\ (\ref{coeff}), (\ref{small_self}), and (\ref{num_den}),
the {\em f}-spin correlation functions contain the whole temperature
dependence of the electronic subsystem (\ref{h_*}).

\section{Results} 
\label{sec:results} 
We have evaluated our theory for a film with simple cubic (s.c.)
structure consisting of $n$ layers parallel to the (100)-plane of the
crystal. The electron hopping and the Heisenberg exchange integrals
shall be restricted within a tight-binding approximation to nearest
neighbour coupling,
\begin{eqnarray}
  \label{tb_hopp}
  T^{\alpha\beta}_{ij} & = & \delta^{\alpha\beta}_{i,j+\Delta}
  T^{\alpha\alpha} + \delta^{\alpha,\beta\pm1}_{ij} T^{\alpha\beta},\\
  \label{tb_exch}
  J^{\alpha\beta}_{ij} & = & \delta^{\alpha\beta}_{i,j+\Delta}
  J^{\alpha\alpha} + \delta^{\alpha,\beta\pm1}_{ij} J^{\alpha\beta},
\end{eqnarray}
respectively. Here $\Delta$ denotes the relative positions of nearest 
neighbours, both within the same layer, for s.c.(100):
$\Delta=(0,1),(0,\bar{1}),(1,0),(\bar{1},0)$. $T^{\alpha\beta}$ is
the hopping between the layers $\alpha$ and $\beta=\alpha\pm 1$ and 
$J^{\alpha\alpha}$ is the exchange interaction within the layer
$\alpha$. For the following discussion, furthermore the hopping
integrals and the exchange interaction have been assumed to be uniform
within the whole film, 
\begin{equation}
  \label{uniform}
  T^{\alpha\alpha}=T^{\alpha\beta} \equiv T, \quad 
  J^{\alpha\alpha}=J^{\alpha\beta} \equiv J_{ff},
\end{equation}
where $J_{ff}$ should not be mixed up with the {\em s-f} exchange
interaction $J$. For explicit values we choose $T=-0.1$eV,
$J_{ff}=0.01$eV,
and the single-ion anisotropy $D_0 / J_{ff} = 0.01$.

\subsection{{\em f}-spin correlation functions}

Before we can start calculating the electronic excitation spectra, we
first have to evaluate the local-moment system considered in 
Sec.\ \ref{subsec:lms}. 
\begin{figure}[htbp]
  \centerline{\epsfxsize=0.83 \linewidth \epsffile{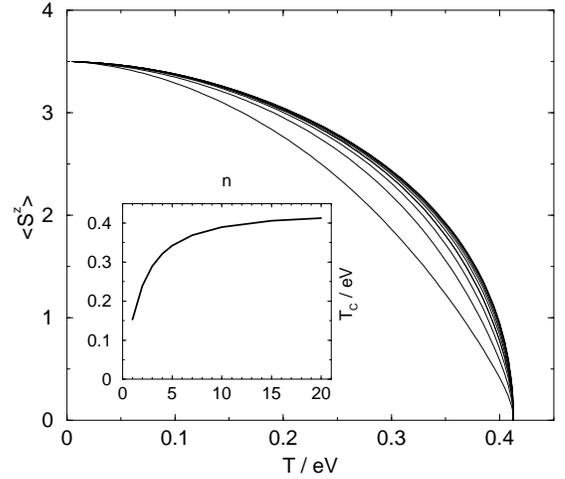}}
  \caption{Layer-dependent magnetizations 
  $\langle S^z_{\alpha} \rangle$ of a 20-layer s.c.(100) film for the
  layers $\alpha=1,2,\ldots,10$. The magnetizations increase from the
  surface layer towards the center of the film. {\bf Inset:}
  Dependence of the Curie temperature $T_C$ on the film thickness $n$ of 
  s.c.-(100) films.}
  \label{fig:1}
\end{figure}
\begin{figure}[htbp]
  \centerline{\epsfxsize=0.83 \linewidth \epsffile{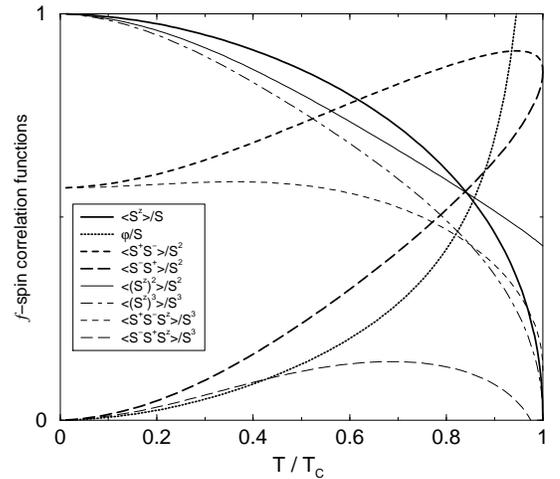}}
  \caption{The different {\em f}-spin correlation functions from
  Eqs. (\ref{coeff}) calculated for the centre layers ($\alpha=10,11$)
  of the 20-layer-film from Fig.\ \ref{fig:1}.}
  \label{fig:2}
\end{figure}
For example, Fig.\ \ref{fig:1} shows the layer-dependent
magnetizations of a 20-layer s.c.(100) film as a function of
temperature. As for all other film thicknesses the layer-dependent
magnetizations $\langle S^z_{\alpha} \rangle$ increase from the surface
layers ($\alpha=1,20$) towards the center layers ($\alpha=10,11$) of the 
film. The inset of Fig.\ \ref{fig:1} displays dependence of 
the Curie temperature on the film thickness $n$.
The complete set of {\em f}-spin correlation functions according to
Eqs.\ (\ref{s_z}), (\ref{varphi}), (\ref{rel_coeff}),
and (\ref{spin_op_rel}) calculated for the center layer of the 20-layer
film can be seen in Fig.\ \ref{fig:2}. 

\begin{figure}[tbp]
  \centerline{\epsfxsize=0.55 \linewidth \epsffile{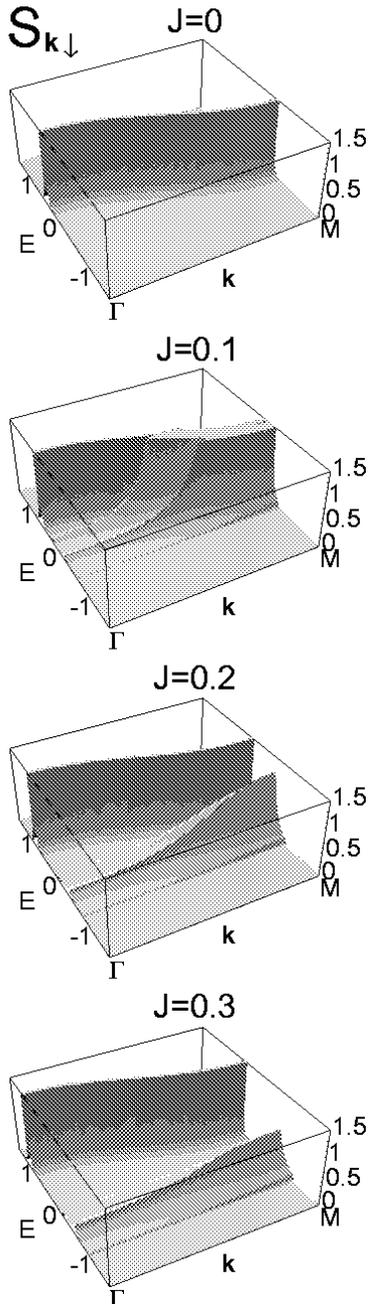}}
  \caption{Spectral density $S_{{\bf k}\downarrow}(E)$ of a
    s.c.-(100)-monolayer for ferromagnetic saturation of the
    local-moment system ($T=0$, $\langle S_z \rangle = S$)
    as a function of energy and wavevector from
    $\Gamma=(0,0)$ to ${\rm M}=(\pi,\pi)$ for $T=0$ 
    and different values of the {\em s-f} exchange interaction $J$.}
  \label{fig:3}
\end{figure}

\subsection{The temperature-dependent electronic structure}
\label{subsec:tdes}
We discuss our results in terms of the spectral density
$S^{\alpha\alpha}_{{\bf k}\sigma}(E)$, defined in Eq.\ (\ref{spectral}),
and the local quasiparticle density of states, Eq.\ (\ref{ldos}).
We start our discussion of the temperature dependent electronic
bandstructure with a special limiting case which gives us an insight
into the underlying physics of the problem.
The special limiting case of ferromagnetic saturation, $T=0$, 
and empty conduction band, $n=0$, is exactly solvable, both, for the bulk
material \cite{AE82,ND85,NMJR96} and for film geometries \cite{SMN97}.
The limiting case, therefore, provides a good testing ground for the 
theory for finite temperatures presented in Sec.\ \ref{subsec:elsub}. 

It turns out that for $T=0$ 
the $\sigma=\uparrow$-spectrum is rather simple, since a
$\uparrow$-electron has no chance to exchange its spin with the
ferromagnetically saturated localized {\em f}-spin system. The
quasiparticle bandstructure is therefore identical to the free Bloch
dispersion, only rigidly shifted by a constant energy amount of
$-\frac{1}{2}JS$, due to the first Ising-like term in the 
{\em s-f} Hamiltonian (\ref{h_sf}).

Fig.\ \ref{fig:3} shows the spin-$\downarrow$ spectral density of a 
s.c.-(100)-monolayer for the special case of ferromagnetic saturation,
$T=0$, for different {\em s-f} interactions $J$. For $J=0$ the spectral
density represents a $\delta$-function located at the point of the free
two-dimensional Bloch dispersion. For small {\em s-f} exchange coupling,
$J>0$, a slight deformation of the original Bloch dispersion sets in and 
the quasiparticle peaks get a finite width indicating a finite lifetime.
For intermediate and strong couplings the spectral density splits into
two parts corresponding to two different spin exchange processes between 
the excited spin-$\downarrow$ electron and the localized {\em f}-spin
system. The higher energetic part of the spectrum represents a
polarization of the immediate spin neighbourhood of the electron due to
a repeated emission and reabsorption of magnons. The result is a
polaron-like quasiparticle called the {\em magnetic polaron}. The
low-energetic part of the spectrum is a scattering band which
corresponds to the simple emission of a magnon by the spin-$\downarrow$
electron, which is necessarily connected with a spin-flip of the 
electron \cite{SMN97}.

\begin{figure}[tbp]
  \centerline{\epsfxsize=0.9 \linewidth \epsffile{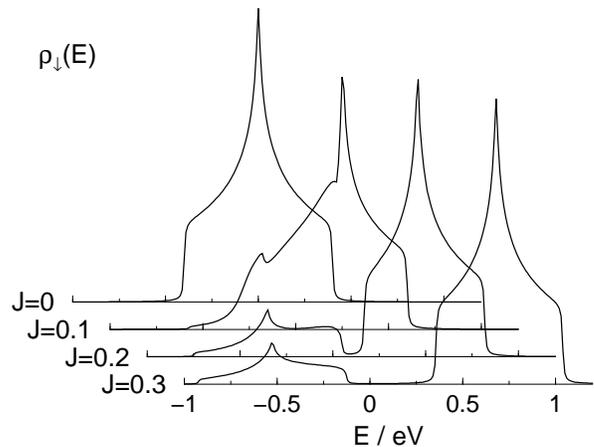}}
  \caption{Density of states $\rho_{\downarrow}(E)$ of a
    s.c.-(100)-monolayer for $T=0$ and
    for different {\em s-f} interactions $J$.}
  \label{fig:4}
\end{figure}

From the spectral density of Fig.\ \ref{fig:3}
we get, using Eq.\ (\ref{ldos}), the local
quasiparticle density of states $\rho_{\downarrow}(E)$ of a monolayer,
displayed in Fig.\ \ref{fig:4}. Here we see that the splitting of the
spectral density discussed above transfers itself to the quasiparticle
density of states as a gap for $J\gtrapprox0.2$. 
As for the spectral density, the density 
of states of the spin-$\uparrow$ electron is only rigidly shifted and
therefore not displayed.

\begin{figure}[bhtp]
  \centerline{\epsfxsize=0.87 \linewidth \epsffile{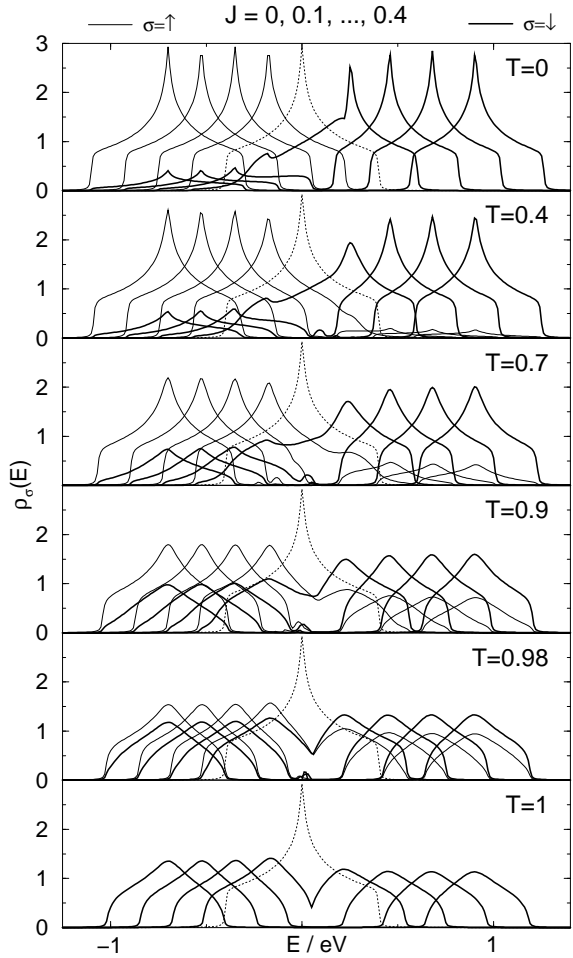}}
  \caption{Density of states $\rho_{\sigma}(E)$ of a
    s.c.-(100)-monolayer for different {\em s-f} interactions $J$ and
    different temperatures $T$ (in units of $T_C$). The dotted lines
    represent the case of $J=0$ where there is no distinction between
    spin-$\downarrow$ and spin-$\uparrow$ electron, 
    $\rho_{\downarrow}(E) = \rho_{\uparrow}(E)$. }
  \label{fig:5}
\end{figure}

However, this does not hold any longer for finite
temperatures, $T>0$.
Fig.\ \ref{fig:5} exhibits the density of states of
a s.c.-(100)-monolayer for different {\em s-f} interactions and
different temperatures. The dotted lines represent the case of
vanishing {\em s-f} exchange, $J=0$, where spin-$\downarrow$ and
spin-$\uparrow$ spectra are equal. Since the electrons are not
coupled to the local-moment system, we also have no temperature dependence. 
For finite {\em s-f} interaction we see from Fig.\ \ref{fig:4} that in
the spin-$\downarrow$ density of states
spectral weight is transferred from the high-energetic polaron peak to
the low-energetic scattering peak. To explain this effect we have to
consider the elementary processes which build the spectrum. The
low-energetic scattering peak of the spin-$\downarrow$ electron consists 
of two elementary processes. 

\begin{figure}[thbp]
  \centerline{\epsfxsize=0.8 \linewidth \epsffile{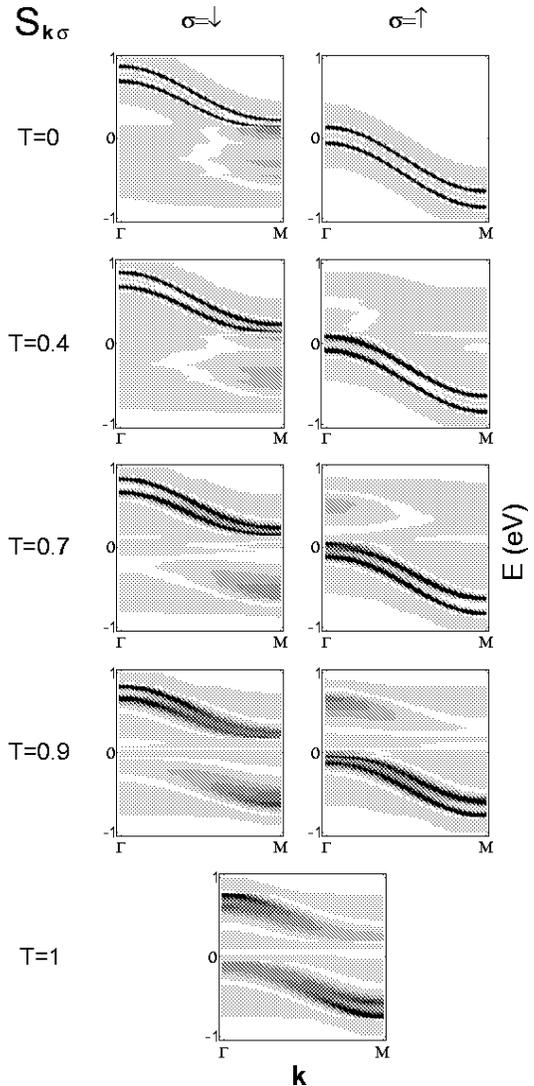}}
  \caption{Spectral density, 
  $S_{{\bf k}\sigma}(E)=S^{11}_{{\bf k}\sigma}(E)=
   S^{22}_{{\bf k}\sigma}(E)$, of a s.c.-(100) double-layer as a
   function of energy and wavevector from $\Gamma=(0,0)$ to 
   $M=(\pi,\pi)$ for $J=0.2$eV and different temperatures $T$ (in units of 
   $T_C$).}
  \label{fig:6}
\end{figure}

Because of finite deviation of the 
{\em f}-spin system from saturation for 
$T>0$, the $\downarrow$-electron has a finite
probability of entering the local frame as spin-$\uparrow$
electron. This probability is zero for $\langle S_z \rangle = S$ ($T=0$) 
and increases with increasing temperature. On the other hand, the
spin-$\downarrow$ electron can
first emit a magnon and by that process reverse its spin, becoming a
spin-$\uparrow$ electron in the external frame of coordinates. 
The spectral weight produced by the first elementary process reduces the 
spectral weight of the high-energetic polaron peak therefore shifting
spectral weight from the high-energetic polaron peaks towards the
low-energetic scattering peak.

For the spin-$\uparrow$ electron we see from Fig.\ \ref{fig:5} that for
finite temperatures an additional peak rises at the high-energetic side of
the spectra with increasing temperature. We can explain this effect by
the spin-$\uparrow$ electron absorbing a magnon and subsequently, as
spin-$\downarrow$ electron forming a polaron. Here the magnon absorption 
by a spin-$\uparrow$ electron is
equivalent to the magnon emission by a spin-$\downarrow$ electron. In
the case of ferromagnetic saturation the system does not contain any
magnons, which is the reason why there is no scattering peak in the
spin-$\uparrow$ spectrum at $T=0$. As a result of the shifting of
spectral weights towards lower energies for the spin-$\downarrow$
electron and towards higher energies for the spin-$\uparrow$ electron
the densities of states for the two spin directions approach each other
with increasing temperature. 

\begin{figure}[htbp]
  \centerline{\epsfxsize=0.95 \linewidth \epsffile{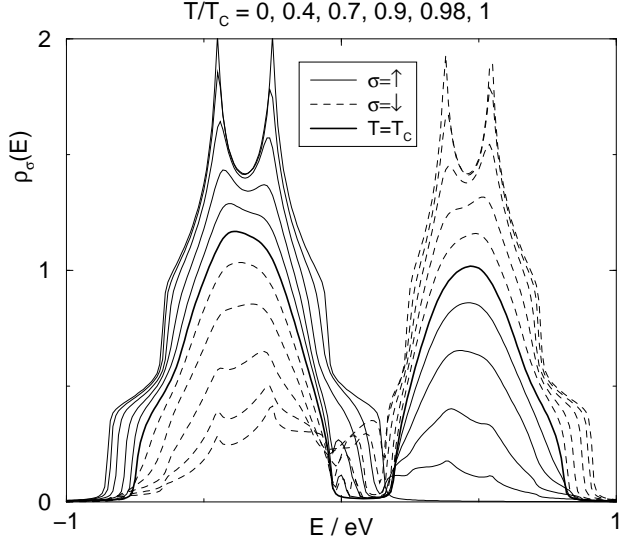}}
  \caption{Density of states,
    $\rho_{\sigma}(E)=\rho^1_{\sigma}(E)=\rho^2_{\sigma}(E)$, of a
    s.c.-(100) double-layer for $J=0.2$ and
    different temperatures $T$ (in units of $T_C$). The curves for $T=0$ are
    furthest away from the paramagnetic solution, $T=T_C$ ({\bf fat} line).}
  \label{fig:7}
\end{figure}

In the limiting case of $T \rightarrow T_C$ the system has eventually
lost its ability to distinct between the two possible spin directions
of the test electron because of the loss of magnetization of the 
underlying local-moment system, $\langle S^z \rangle \rightarrow 0$. 
Hence as for the case of vanishing 
{\em s-f} interaction for $T=T_C$ the density of states of states of the 
spin-$\downarrow$ electron equals that of the spin-$\uparrow$ electron.
Another feature which can be seen from Fig.\ \ref{fig:5} is that the
positions of the four quasiparticle subbands, two for each spin
direction, do not change with temperature. 

To further discuss the temperature effects we present with
Figs. \ref{fig:6} and \ref{fig:7} the spectral density and the local
density of states, respectively, of a s.c.-(100) double-layer ($n=2$) for
$J=0.2$ and different temperatures. Again we see that the spectra
for the two spin directions approach each other for $T\rightarrow T_C$.
Another feature which can already be observed in Fig.\ \ref{fig:5} is
that the increase of temperature results in the narrowing of the
subbands. For the case of intermediate coupling, $J=0.2$, according to
Figs.\ \ref{fig:6} and \ref{fig:7} this band narrowing results in the
opening of a gap between the scattering and the polaron band with
increasing temperature. 

This temperature enhanced band splitting has
already been found for the three-dimensional case \cite{NMJR96}. 
It can be explained for the spin-$\uparrow$ electron by the fact that
for propagating in its own low-energetic subband it needs to find an
appropriate lattice site. 
\begin{figure}[htbp]
  \epsfxsize=\linewidth \epsffile{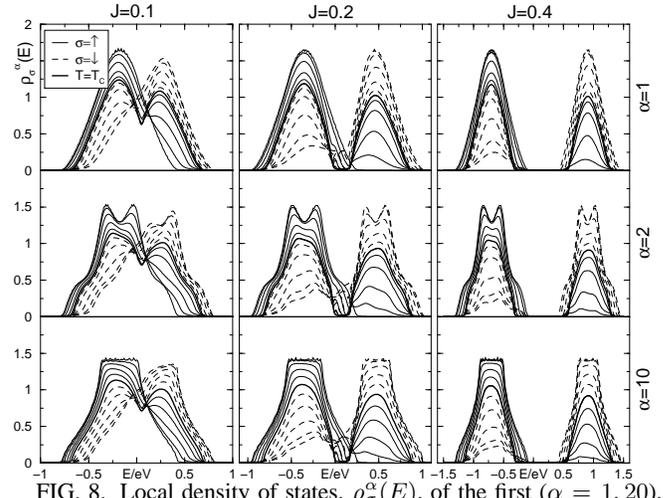}
  \caption{Local density of states, $\rho_{\sigma}^{\alpha}(E)$, of the
    first ($\alpha=1,20$), second ($\alpha=2,19$), and center ($\alpha=10,11$)
    layer of a 20-layer s.c.-(100) film for different {\em s-f}
    interactions $J$ and different temperatures 
    $T/T_C = 0,0.4,0.7,0.9,0.98,1$. The curves for $T=0$ are
    furthest away from the paramagnetic solution, $T=T_C$ ({\bf fat} line).}
  \label{fig:8}
\end{figure}
In the case of ferromagnetic saturation there
is no restriction for the propagation of the spin-$\uparrow$ electrons
since spin-flip processes are impossible. With increasing temperature
there is an increasing deviation of the local moments resulting in the
possible magnon absorption by the spin-$\uparrow$ electron and
subsequent changing to the higher energetic polaron subband. 
Hence, the
spin-$\uparrow$ electron, to propagate in its own subband needs to move
further distances with increasing temperature resulting in a reduced
effective hopping and in a decreased bandwidth.
\begin{figure}[htbp]
  \epsfxsize=\linewidth \epsffile{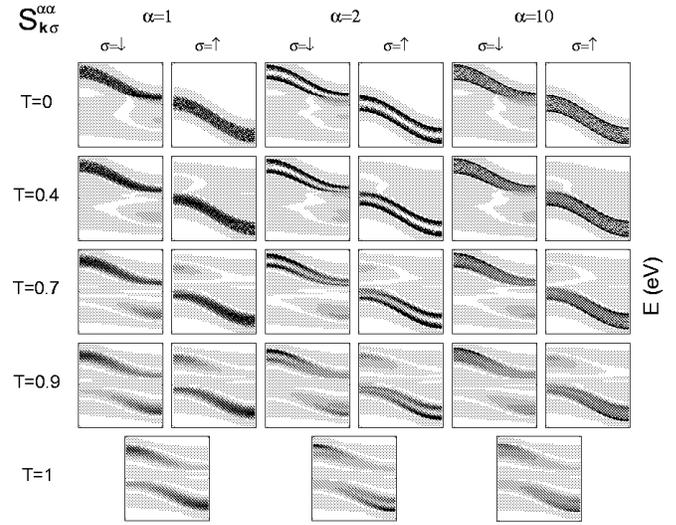}
  \caption{Spectral density, $S^{\alpha\alpha}_{{\bf k}\sigma}$, as a
    function of wavevector ${\bf k}$ from $\Gamma=(0,0)$ to 
    ${\rm M}=(\pi,\pi)$ (horizontal axes) and energy $E$ from -1eV to
    1eV (vertical axes) of the
    first ($\alpha=1,20$), second ($\alpha=2,19$), and center ($\alpha=10,11$)
    layer of a 20-layer s.c.-(100) film for $J=0.2$eV and different
    temperatures (in units of $T_C$).}
  \label{fig:9}
\end{figure}

In addition to the discussed temperature effects, Figs.\ \ref{fig:6} and 
\ref{fig:7} exhibit a typical two-peak structure which is caused by the
coupling of the two layers. This two-peak structure is replaced in the
case of a film consisting of $n$ equivalent layers by an $n$-peak structure.
Generally, the spectra of the discussed local-moment films are
characterised by an interplay between correlation ($J$), temperature
effects, and geometry of the film. Figs.\ \ref{fig:8} and \ref{fig:9}
display results for the local density of states and the layer-dependent
spectral density of a 20-layer s.c.-(100)-film. Additionally to the
dependence on the {\em s-f} exchange interaction and the temperature
dependence we notice that the spectral density and the density of 
states show a typical layer-dependence due to the broken translational
symmetry at the surfaces of the film
\cite{SMN97}. For the centre layers ($\alpha=10,11$) of the
20-layer-film we see from Fig.\ \ref{fig:8} that the local density of states
of the spin-$\uparrow$ electron at $T=0$ has already become pretty
similar to the well-known tight-binding density of states of the
three-dimensional s.c. lattice whereas the density of states of the
surface layers ($\alpha=1,20$) exhibits the characteristic 
semi-elliptic profile.

\section{Summary} 
\label{sec:summary} 
We have investigated the electronic quasiparticle bandstructure of a
ferromagnetic semiconductor film. A single test electron ({\em s}-band)
is coupled by an intra-atomic {\em s-f} inter-band exchange to a system
of localized {\em 4f}-moments. That may be regarded as a proper model
description for Euo and EuS. Our approach uses a moment-conserving
decoupling procedure for suitable defined Green functions. The fact that 
our theory evolves continuously from the exactly solvable limiting case
of ferromagnetic saturation \cite{SMN97} gives it a certain trustworthiness.

The exchange coupling of the conduction electron to the local-moment
system gives rise to a correlation induced splitting of the
quasiparticle spectra. A polaron part may be interpreted
as a repeated emission and reabsorption of magnons by the conduction
electron resulting in a new quasiparticle, the magnetic polaron. A
rather broad scattering peak is due to a simple magnon emission
or absorption by the conduction electron. This pronounced splitting
depends on the actual value of the exchange interaction
$J$. For small values of $J$ only a renormalization of the one-electron
energy occurs resulting in a deformation of the free Bloch dispersion.
For higher values of $J$, the mentioned splitting of the spectra into
polaron part and scattering part sets in.

We intend to apply the presented model to study the temperature
dependent electronic structure of EuO and EuS films. Therefore the
electronic part of the Hamiltonian (\ref{h_s}) and the {\em s-f} exchange
interaction (\ref{h_sf_orig}) have to be modified to include the multiband 
aspect of real substances. This will be done by substituting the
tight-binding bandstructure (\ref{tb_hopp}) by a realistic one
taken from a bandstructure calculation.

Another highly interesting field which we want to use our theory for is
the evaluation of the temperature dependence of surface states. In a
previous paper we have calculated surface states for the special case of 
$T=0$ and $n=0$ by modifying the hopping in the vicinity of the surface
\cite{SMN98}. The extension of these calculations to finite
temperatures promises to give an understanding of recent experimental results
concerning the temperature dependence of electronic states on
surfaces of rare earths.

\section*{Acknowledgement} One of the authors (R. S.) acknowledges the
support by the German National 
Merit Foundation. The support by the Sonderforschungsbereich 290
(''Metallische d\"unne Filme: Struktur, Magnetismus und elektronische
Eigenschaften``) is gratefully acknowledged.



\end{document}